\documentclass[preprint]{aastex}
\usepackage[twocolumn]{emulateapj5}

\newcommand{\Tin}{T_{\rm in}}
\newcommand{\Rin}{R_{\rm in}}
\newcommand{\rg}{r_{\rm g}}
\newcommand{\Tcol}{T_{\rm col}}
\newcommand{\Teff}{T_{\rm eff}}

\slugcomment{2000/11/23~~$ApJL$ accepted }

\shorttitle{Watarai et al.}
\shortauthors{Slim Disk Model for ULXs}

\begin{document}
\title{Slim Disk Model for Ultra-Luminous X-Ray Sources }

\author{Ken-ya Watarai\altaffilmark{1,3}, 
        Tsunefumi Mizuno\altaffilmark{2},
        Shin Mineshige\altaffilmark{1}}

\altaffiltext{1}{Department of Astronomy, Graduate School of Science,
Kyoto University}
\altaffiltext{2}{Department of Physics, Faculty of Science, 
Hiroshima University}
\altaffiltext{3}{$E$-$mail:$ watarai@kusastro.kyoto-u.ac.jp}

\begin{abstract}
The Ultra Luminous X-ray Sources (ULXs) are unique in exhibiting
moderately bright X-ray luminosities,
$L_{\rm x} \sim 10^{38-40} {\rm erg~s^{-1}}$, and
relatively high blackbody temperatures, $\Tin \sim 1.0-2.0 {\rm keV}$.
 From the constraint that $L_{\rm x}$ cannot exceed
the Eddington luminosity, $L_{\rm E}$,
we require relatively high black-hole masses, $M\sim 10-100 M_\odot$,
however, for such large masses the standard disk theory predicts
lower blackbody temperatures, $\Tin < 1.0$ keV.
To understand a cause of this puzzling fact, 
we carefully calculate the accretion flow structure 
shining at $\sim L_{\rm E}$,
fully taking into account the advective energy transport 
in the optically thick regime and the transonic nature of the flow.
Our calculations show that at high accretion rate 
($\dot M \ga 30~L_{\rm E}/c^2$)
an apparently compact region with a size of 
$\Rin \simeq (1-3)\rg$ (with $\rg$ being Schwarzschild radius) 
is shining with a blackbody temperature of
$\Tin \simeq 1.8 (M/10M_\odot)^{-1/4}$ keV
even for the case of a non-rotating black hole.  
Further, $\Rin$ decreases as $\dot M$ increases, 
on the contrary to the canonical belief that the inner edge of the disk 
is fixed at the radius of the marginally stable last circular orbit.
Accordingly, the loci of a constant black-hole mass 
on the {\lq\lq}H-R diagram" (representing the relation between 
$L_{\rm x}$ and $\Tin$ both on the logarithmic scales) 
are not straight but bent towards the lower $M$ direction
in the frame of the standard-disk relation.

We also plot the ASCA data of some ULXs on the same H-R diagram, 
finding that they all fall on the regions with
relatively high masses, $M \sim 10-30M_{\odot}$, and
high accretion rates, ${\dot M}\ga 10~L_{\rm E}/c^2$.
Interestingly, IC342 source 1, in particular, was observed to
move along the constant-$M$ line (not constant $\Rin$ line)
in our simulations.
This provides a firm evidence that at least some ULXs
are shining at $\ga L_{\rm E}$, and containing black holes
with $M \simeq 10-100 M_\odot$.
\end{abstract}

\section{Introduction}

Already in the late 1980's,
$Einstein$ satellite had found
luminous X-ray sources with X-ray luminosities
$L_{\rm x} \sim 10^{38-40}~{\rm erg~s^{-1}}$
in nearby spiral galaxies (Fabbiano.\ 1989).
They locate in the off-nucleus region of the host galaxy,
hence cannot be an active-galactic nucleus (AGN) of low-luminosity.
Only a small number of them are optically identified as young
supernova remnants (e.g., the most luminous source in NGC~6946;
Schlegel\ 1994), whereas we lack secure optical identifications
for most of them.
At present day, these luminous,
non-AGN and non-supernova-remnant
sources are called
{\lq\lq}Ultra Luminous X-Ray Sources (ULXs)"
(see Makishima et al.\ 2000 and reference therein).
Apparently, their X-ray luminosities greatly exceed the Eddington
limit for a neutron star, $\sim 2\times 10^{38}$ erg~s$^{-1}$,
by a couple of orders of magnitude.
ULXs are thus more like black-hole binaries (BHBs),
although we need spectral information to reveal their 
emission mechanism.

Recently, much progress has been made on the study of ULXs.
Thanks to the fine spectral and imaging capability of $ASCA$,
high-quality spectra have been accumulated for numbers of ULXs,
including NGC~1313 source~A and B (Petre et al.\ 1994),
M81~X-6 (Uno 1996),
two ULXs in IC~342 (Okada et al.\ 1998), and so on.
Especially, rather extensive study has been performed
by Makishima et al.\ (2000) and Mizuno et al. (2000).

According to their study,
the spectra of most ULXs in nearby galaxies can be well fitted
with multi-color disk (MCD) model (Mitsuda et al.\ 1984),
reinforcing its BHB interpretation.
However, obtained disk temperatures, typically ranging 1.0--2.0~keV,
somewhat contradict the BHB hypothesis.
Theoretically, they are
too high to be achieved by a standard-type disk around such a
high-mass, Schwarzschild (non-rotating) black hole.
They are also higher than those observed from
well-studied Galactic/Magellanic
BHBs (e.g., Tanaka \& Lewin\ 1995).
The observed high disk temperatures are hence
the most severe problem regarding ULXs.
To cope with this issue, some authors discussed the possibility
of Kerr BH,
although their arguments still remain rather qualitative.
We need both the detailed calculation of spectra from BHBs
and the careful comparison with the observations.

In this $Letter$ we consider an important process
which has not been fully appreciated in the study of ULXs.
The accretion-disk theory predicts that for high luminosity,
comparable to $L_{\rm E}$,
advective energy transport dominates over energy loss by radiation.
Abramowicz et al. (1988) were first to construct such a disk model,
currently known as the slim disk model
(see also Szuszkiewicz et al. 1996;
Wang et al. 1999).
More recently,
Watarai et al. (2000) and Mineshige et al. (2000) 
discussed the distinctive spectral features 
of the slim disks in connection with BHBs
and narrow-line Seyfert 1 galaxies (NLS1s), respectively
(see also Fukue 2000).
Following the same line, we calculate the structure
and emission properties of high-luminosity accretion disks
and demonstrate that the X-ray spectral properties of ULXs are 
indeed well understood by the slim disk model.
In section 2 we give basic equations with advection, 
where we stress the necessity of solving flow structure 
with careful consideration of its transonic nature.
We then show in section 3 the results of model fitting 
and compare the theory with ULXs observed by $ASCA$.
In section 4, we argue that ULXs do really undergo super-critical
accretion.  The final section is devoted to conclusions.
Throughout the present study, we normalize
$\dot M$ by $L_{\rm E}/c^2$ (with $c$ being speed of light);
i.e., the normalized mass-flow rate is
 $ {\dot m} \equiv \dot M/(L_{\rm E}/c^2)
     \simeq \dot M/
                 (1.4\times 10^{17}(M/M_\odot)~{\rm g~s}^{-1})$.

\section{Basic Equations and Numerical Methods}

The calculation code of the transonic accretion flow was first
made by R. Matsumoto (see Matsumoto et al. 1984)
and then modified by F. Honma and M. Takeuchi so as to incorporate
the advected energy term.
Basic equations used here are the same as those in
the previous paper (Watarai et al. 2000).
We assume:
(1) the disk is steady ($\partial /\partial t = 0$);
(2) the disk is axisymmetric;
(3) the Pseudo-Newtonian approximation (Paczy\'nsky \& Wiita 1980)
  is adopted for a gravitational potential, $\psi = -GM/(r-r_{\rm g})$
  with $r_{\rm g}$ being the Schwarzschild radius;
(4) the vertical disk structure is integrated.
The energy equation is symbolically written as
   $ Q_{\rm adv}^- = Q_{\rm vis}^+ - Q_{\rm rad}^-$, 
where the two terms on the R.H.S. represent viscous heating
and radiative cooling, while the term on the L.H.S. is
advection term
(see Kato et al. 1988 for explicit expression).

We solved the full-set of the basic differential equations with
semi-implicit method for appropriates boundary conditions.
Black-hole accretion flow is transonic; it is subsonic far outside,
whereas it is supersonic near the black hole.  Thus,
the solution should satisfy the regularity condition 
at the transonic point.
The flow structure is calculated 
from the outer edge located at $r_{\rm out}=10^4 r_{\rm g}$
to $\rg$ through the transonic point (barely inside $\sim 3~\rg$).
At the outer boundary, we impose the same physical
quantities as those of the standard disk model, while at the inner
boundary we assume torque-free conditions.
For estimating the effective temperature, we assume   
the radiative cooling energy with black body radiation.
For moderately large viscosity parameter, $\alpha > 0.1$,
this assumption may break down (Beloborodov 1998),
thus we assume relatively small $\alpha = 0.01$
although spectra do not sensitively depend on $\alpha$
(Watarai et al. 2000, in preparation;
see also discussion in Mineshige et al. 2000).
Then, the local effective temperature is 
$T_{\rm eff} =(Q_{\rm rad}/2 \sigma)^{1/4}$.
The Compton scattering within the disk
will give rise to higher color temperature, $\Tcol > \Teff$.
Such effects will be parameterized later.
We assume the face on disk ($i=0$).

\section{Model Spectral Fitting}







The MCD model has been very successful in analyzing 
soft X-ray spectra of low-mass X-ray binaries. 
However, it is valid only for the standard-type disks (with $L \ll L_{\rm E}$)
in steady state, since it assumes that the temperature profile of the disk is 
$T \propto r^{-3/4}$. Mineshige et al. (1994) generalized the MCD model
by setting the temperature gradient to be $T \propto r^{-p}$
with $p$ being a fitting parameter.
We use this {\lq\lq}$p$-free model," 
which is quite useful for discriminating slim disks showing $p=0.5$
from the standard one with $p=0.75$ (Watarai et al. 2000). 
Our fitting results of theoretically calculated spectra
are summarized in Table 1. 
%
%

Figure 1 displays the X-ray {\lq\lq}H-R diagram" 
representing the relationship between X-ray luminosities and 
X-ray temperatures.  In this figure,
we plot the loci of constant $M$ and constant $\dot M$
based on the standard disk model (with dotted lines)
and those based on our calculations (with solid lines),
As was stated in Makishima et al. (2000), bolometric luminosity of 
the accretion disk is determined by the observational 
physical quantities,
\begin{eqnarray}
 L_{\rm bol} &=& 
    7.2 \times 10^{38} \left(\frac{\xi}{0.41}\right)^{-2} 
               \left(\frac{\Tcol/\Teff}{1.7}\right)^{-4} \nonumber \\
    & & \times \left(\frac{\Rin}{3 \rg}\right)^2 
               \left(\frac{M}{10M_{\odot}}\right)^2
               \left(\frac{\Tin}{\rm keV}\right)^4 {\rm erg~s^{-1}}.
\end{eqnarray}
where $\xi$ is a correction factor needed for the MCD model
 (Kubota at al. 1998), since the MCD model tends to over-estimate
the radius of the inner edge due to the negligence of the boundary
term in the $\Teff$ expression, and
$\Tcol/\Teff$ is a spectral hardening factor (Shimura \& Takahara 1995),
which relates the observed color temperature ($\Tcol$) 
and effective temperature ($\Teff$) of the disk.
[Spectral hardening occurs because of internal Compton scattering,
see Czerny \& Elvis 1987; Ross, Fabian, \& Mineshige 1992.]
Note that normalized luminosity,
$\eta \equiv L/L_{\rm E}$, is mass-independent,
as long as $\dot M$ ($\propto L$) is measured 
in a unit of $L_{\rm E}/c^2$.
When plotting our results in figure 1,
we also include the corrections corresponding
to $\xi$ and $\Tcol/\Teff$ above 
so that we can directly compared with the observational results 
using MCD approximation.

Clearly, our results roughly coincide with 
those of the standard disks in the low-$\dot M$ (lower-left) regions.
A small discrepancy 
is due to the different adopted potentials:
we used the pseudo-Newtonian potential,
while the standard disk is based on the Newtonian one.
In high-$\dot M$ regions, in contrast, our calculation results 
systematically shift towards the lower $M$ direction in the frame of the
standard-disk relation for given $\dot M$.
This is due to the apparent shift of the inner boundary
in transonic flows with high $\dot M$.
The behavior of the two models on the X-ray H-R diagram
is thus qualitatively distinct at high $L$.

In figure 1, we also plot
$ASCA$ observation results of ULXs and BHBs.
Makishima et al. (2000) analyzed $ASCA$ data of several ULXs and
clarified their unique X-ray observational features at 0.5-- 10keV.
We use the data sample of ULXs given by Mizuno et al. (2000), together with
other samples of galactic sources from Makishima et al. (2000)
 for comparison (see Table 2).  They show that
ULXs tends to gather in the region with high $M$ and $L$,
compared with BHBs in figure 1.

We, here, pay particular attention to the multiple points of
IC342 source 1 in figure 1, which indicate that
this source moves along the constant-$M$ lines 
not of the standard disk but of the slim disk.
Obviously, black-hole masses cannot change, while
$\dot M$ can on short timescales, say, $< 1$ s.
That is, the slim disk is more relevant here.
Other objects, such as $M81$ X-8 and $NGC1313$ source B,
also move along our constant $M$ lines.
To see these more clearly,
we plot in figure 2 the same quantities to those in figure 1
but on the $\Rin$-$\Tin$ diagram.
We can explicitly see in this figure
how $\Rin$ changes as $\dot m$ varies.
These strongly support that we are actually observing
changes in $\Rin$ in accord with variations in $\dot M$,
which cannot be explained in the framework of the standard disk model.


\begin{deluxetable}{rrrr} 
\tabletypesize{\footnotesize}
\tablecolumns{4} 
\tablewidth{0pc} 
\tablecaption{Fitting results of p-free model (0.2-10keV, $M=33M_\odot$)} 
\tablehead{ 
\colhead{$\dot{M}/(L_{\rm E}/c^2)$} & \colhead{$T_{\rm in}$ (keV)}   
& \colhead{$R_{\rm in}$ (km)}    & \colhead{p}}
\startdata 
1    & 0.24(0.2-3.0 keV) & 758.6 & 0.73 \\ 
3    & 0.33(0.2-5.0 keV) & 724.4 & 0.73 \\ 
10   & 0.45              & 691.8 & 0.72 \\ 
33   & 0.66              & 501.2 & 0.67 \\ 
100  & 1.21              & 165.9 & 0.59 \\ 
333  & 1.48              & 117.5 & 0.55 \\ 
1000 & 1.61              & 95.5 & 0.53 \\ 
\enddata 
\end{deluxetable} 


\begin{deluxetable}{rrrrrr} 
\tabletypesize{\footnotesize}
\tablecolumns{8} 
\tablewidth{0pc} 
\tablecaption{Fitting results of ULXs with ASCA (0.5-10keV) } 
\tablehead{ 
\colhead{Source} 
& \colhead{$T_{\rm in}$ (keV)}   
& \colhead{$R_{\rm in}$ (km)} 
& \colhead{$L_{\rm bol}$ ($10^{38}~{\rm erg~s^{-1}}$)} 
& \colhead{}  & \colhead{}}
\startdata 
IC342 source 1 & 1.96{$\pm 0.1$} & 104{$\pm 9$}	  & 147 &  &  \\ 
       	& 1.5{$\pm 0.1$}	& 142{$\pm 18$}	  & 93 &  &  \\ 
	& 1.7{$\pm 0.15$}	& 125{$\pm 21$}	  & 118 &  &  \\ 
	& 1.29{$\pm 0.08$}	& 168{$\pm 20$}	  & 72 &  &  \\ 
	& 1.81{$\pm 0.07$}	& 116{$\pm 8$}	  & 137 &  &  \\ 
NGC1313 source B & 1.47{$\pm 0.08$} & 110{$\pm 12$}	& 50 &  &  \\ 
       	         & 1.07{$\pm 0.07$} & $129_{-16}^{+19}$	& 20 &  &  \\ 
M81 source X6 & 1.59{$\pm 0.09$} & 72{$\pm 7$}	   & 30 &  &  \\ 
       	& 1.29{$\pm 0.13$}	& $89_{-15}^{+20}$ & 20 &  &  \\ 
\enddata 
\end{deluxetable} 


\section{Nature of ULXs: Really Undergoing Super-Critical Accretion?}

The most common interpretation often made
to explain high $\Tin$ and $L$ of ULXs is that
ULXs may contain Kerr black holes (Zhang et al. 1997).
For Kerr holes,
the marginally stable last circular orbit can be located at smaller
radii, down to $\sim 0.5\rg$ in an extreme case, thus yielding small
$\Rin$ and, hence, high $\Tin$.  Also,
the energy conversion efficiency from accretion energy to
radiation energy can be as large as 0.42 at maximum,
thereby generating large $L$.

However, we wish to stress that
non-advective, non-transonic models are self-inconsistent in high 
$\dot M$ regimes,
because advective energy transport dominates over radiative cooling.
We, in fact, see in figures 1 and 2 that
typical accretion rates of ULXs are $\dot m \simeq 30-100$,
for which a transition from the standard-type disk to the slim disk
occurs.
We can roughly express the disk luminosity as a function of $\dot m$
(Watarai et al. 2000),
\begin{equation}
 L({\dot m}) \simeq L_{\rm E} \left\{ 
\begin{array}{lcl}
    2[{1 + \ln({\dot m}/20)}] & \mbox{for} & {\dot m} \geq 20 \\
       {\dot m}/10            & \mbox{for} & {\dot m} < 20.
\end{array} \right.
\end{equation}
That is, we can expect significant advection effects for
$\dot m\ga 20$.
Likewise, we expect significant amount of mass existing
around inside the marginally stable last circular orbit.
%
The similar situation may be realize in different circumstances,
e.g. in the represent magnetic stress (Agol \& Krolik 2000), 
in 2D numerical simulation (Stone et al. 1999), 
although there are no quantitative discuss have been made. 

%

The distinction between the standard disk and the slim disk
is not only quantitative but also qualitative.
More important is that the Kerr model cannot account for 
time changes in $\Rin$.
In short, the peculiar behavior of some ULXs in figures 1 and 2
remain unexplained in the framework of usual disk models.
It might be that there exists a slim disk surrounding a Kerr hole.


We assume the face on disk ($i=0$) and do not consider 
the relativistic effects, except for adopting the pseudo-Newtonian 
potential, in these figures. 
General relativistic (GR) effects (such as gravitational redshift,
Doppler boosting, gravitational focusing due to ray bending)
produce complex effects, depending on inclination angles.
We calculate a slim disk model
with face-on geometry, for which only gravitational redshift works,
finding systematically $larger$ $\Rin$ ($\sim 3 r_{\rm g}$)
by a factor of $\sim 6$
and thus $lower$ $\Tin$ ($\sim 0.12$ keV)
by a factor of $\sim 2.5$ even for $\dot m \ga 100$.
In other words, most of radiation from inside $3 \rg$ vanishes
for this case.
For a face-on disk, therefore, GR effects cause
decrease in $\Tin$ and increase in $\Rin$, keeping roughly constant $L$.
Corrections to remove Compton and GR effects from the data points are,
hence, $\Delta \log\Tin \la +0.4$ and
$\Delta \log\Rin \la -0.8$.

For disks with non-zero inclination angles 
around a Schwarzschild black hole, Doppler boosting enhances
radiation, yielding, at most, a factor of $\sim 1.4$ $increase$
in $\Tin$ compared with face-on disks,
while the total flux kept roughly constant (Sun, Malkan 1989).
We thus expect slight $increase$ in $\Tin$
and $decrease$ in $\Rin$ for a fixed $\Rin^2\Tin^4$
compared with face-on disks.  
Required corrections are
$\Delta\log \Tin \la -0.2$ and
$\Delta\log \Rin \la +0.4$.

To summarize, the inclusion of GR and inclination effects
cause large uncertainties in $\Rin$ and $\Tin$.
It then follows that we cannot obtain a good estimate of
$m$ and $\dot m$ from figure 1,
unless an inclination angle is accurately known.
Nevertheless, our conclusion that ULXs are near critical accretion
does not alter, since the positions of some ULXs in figures 1 and 2 
really move along the qualitatively different path than 
that expected by the standard disk model.
It is interesting to point
that our calculations can well reproduce the observations,
even without considering the GR and inclination effects.
We also understand that
sources such as IC342 source 1, $M81$ X-8, and $NGC1313$ source B
cannot be face-on, for the reason that if so we do not expect $\Rin$ changes,
since radiation from inside $\sim 3~\rg$ will vanish.



\section{Conclusions}


\noindent{1.}
All the ULXs so far observed fall on the regions with
high mass ($M \sim 30M_\odot$) and high accretion rate
($\sim 30 L_{\rm E}/c^2$) on the X-ray H-R diagram (figure 1).

\noindent{2.}
The slim disk model predicts apparent decrease in $\Rin$
as $L$ increases.  This results in a qualitatively different
evolutionary path of a single object on the X-ray H-R diagram, 
compared with that of the standard disk (i.e., constant $\Rin$ line).
ULXs that are observed at several different epochs 
(only IC342 was observed at same epochs) actually
follow the constant-$M$ lines of the slim disk, not those of the
standard disk.
This provides a firm evidence that at least some ULXs are near
critical accretion phase.

\noindent{3.}
It is not always clear whether a central black hole is rotating or not.
What we can conclude is that high luminosity, $\sim L_{\rm E}$, is realized
in ULXs, not depending on the nature of black holes.

\noindent{4.}
In future work,
we need more accurate evaluations of GR effects and
the effects of changing viscosity prescriptions.
Also, more extensive observational studies, particularly
independent estimates of inclination angles, are indispensable 
for fully understanding the nature of ULXs.

\acknowledgments
We would like to thank Dr. K. Makishima for useful
comments and discussions.
This work was supported in part by the Grants-in Aid of the
Ministry of Education, Science, Sports, and Culture of Japan
(10640228, SM).
\newpage
\centerline{
\vbox{\epsfxsize=8.cm
\epsfbox{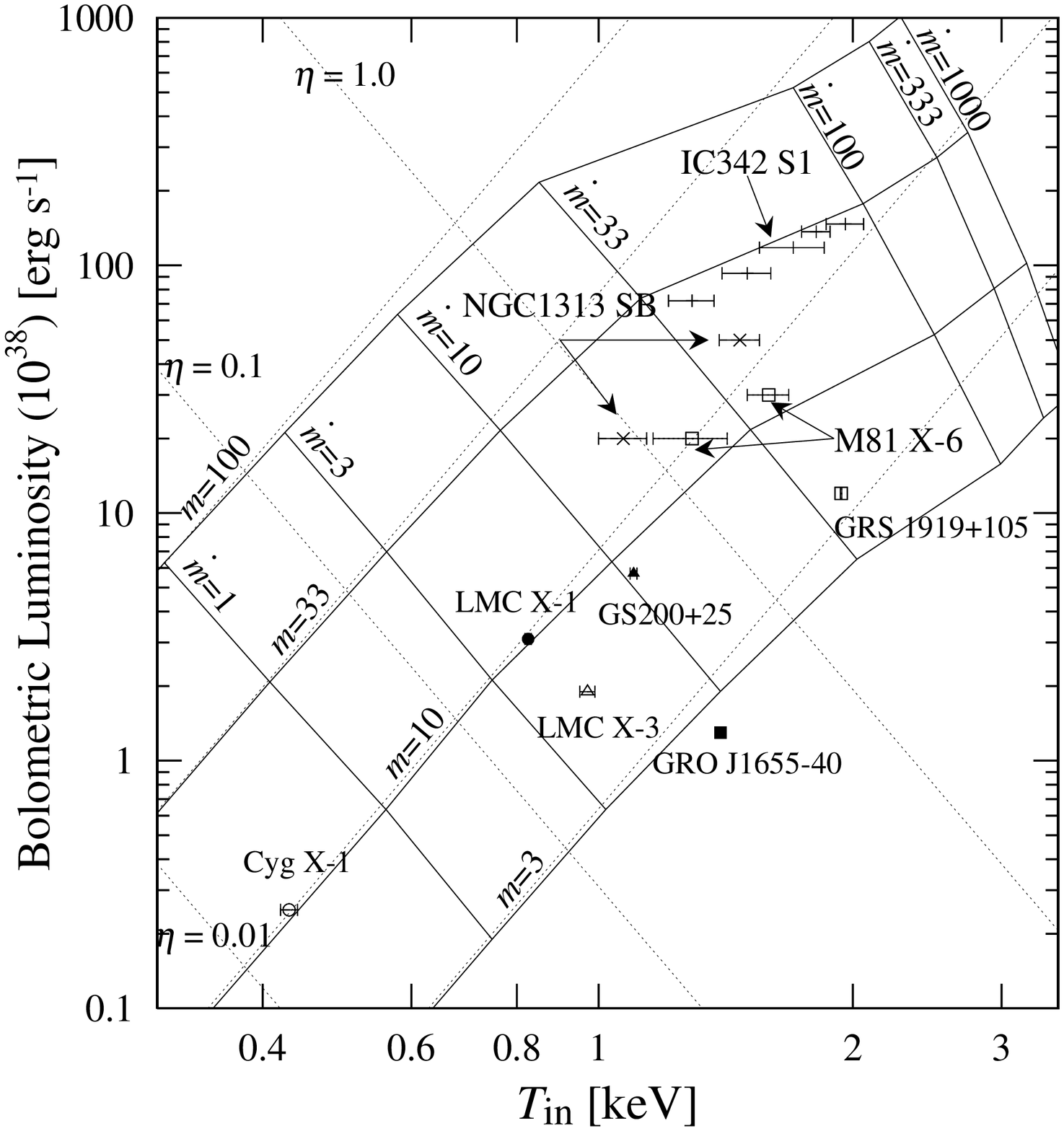}}}
{\small F{\scriptsize IG}.~1.---
X-ray H-R diagram of X-ray sources. 
Solid lines represent the constant $m$ (black-hole mass)
and constant $\dot m$ loci according to our model, while
 dotted lines are the same but based on the 
standard accretion disk (SSD) theory. 
When calculating both lines, the boundary ($\xi$) and Compton ($\Tcol/\Teff$)
effects are both included.  
Other symbols denote the $ASCA$ data of ULXs and BHBs taken
from Mizuno (2000) and Makishima et al. (2000). 
The inclination angle is assumed to be $i$=0 (face on). }

\centerline{
\vbox{\epsfxsize=9.0cm
\epsfbox{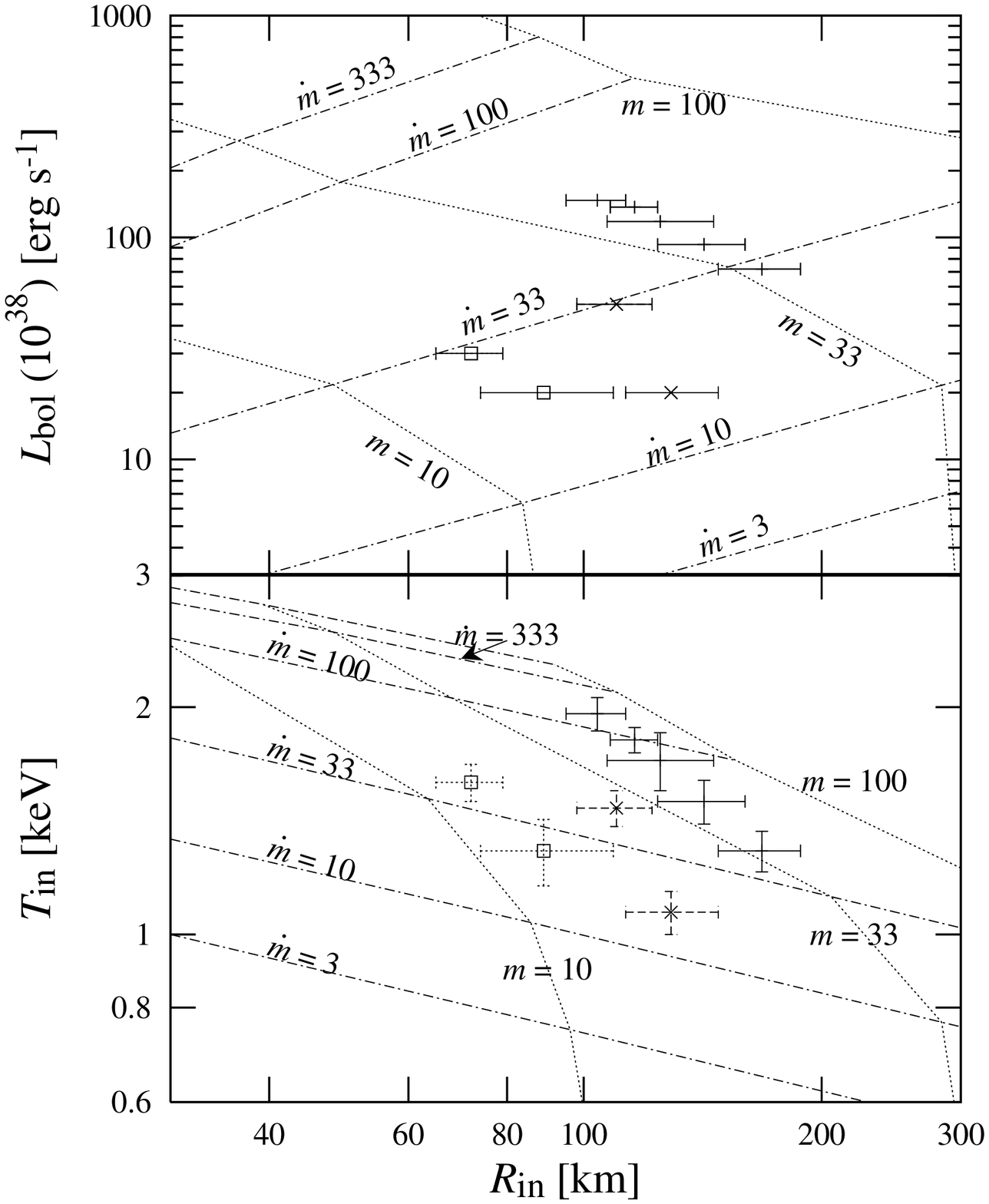}}}
{\small F{\scriptsize IG}.~2.---
The same as figure 1 but on the $\Rin$-$L_{\rm bol}$
(bolometric luminosity), $\Rin$-$\Tin$ diagram.
Dotted and dash-dotted lines represent the constant $m=M/M_{\odot}$ 
and $\dot{m}$, respectively.}
\\

\clearpage





\begin{deluxetable}{rrrr} 
\tablecolumns{4} 
\tablewidth{0pc} 
\tablecaption{Fitting results of p-free model (0.2-10keV) ($M=33M_\odot$)} 
\tablehead{ 
\colhead{$\dot{M}/(L_{\rm E}/c^2)$} & \colhead{$T_{\rm in}$ (keV)}   
& \colhead{$R_{\rm in}$ (km)}    & \colhead{p}}
\startdata 
1    & 0.24(0.2-3.0 keV) & 758.6 & 0.73 \\ 
3    & 0.33(0.2-5.0 keV) & 724.4 & 0.73 \\ 
10   & 0.45              & 691.8 & 0.72 \\ 
33   & 0.66              & 501.2 & 0.67 \\ 
100  & 1.21              & 165.9 & 0.59 \\ 
333  & 1.48              & 117.5 & 0.55 \\ 
1000 & 1.61              & 95.5 & 0.53 \\ 
\enddata 
\end{deluxetable} 

\begin{deluxetable}{rrrrrr} 
\tablecolumns{8} 
\tablewidth{0pc} 
\tablecaption{Fitting results of ULXs with ASCA (0.5-10keV) } 
\tablehead{ 
\colhead{Source} 
& \colhead{$T_{\rm in}$ (keV)}   
& \colhead{$R_{\rm in}$ (km)} 
& \colhead{$L_{\rm bol}$ ($10^{38}~{\rm erg~s^{-1}}$)} 
& \colhead{}  & \colhead{}}
\startdata 
IC342 source 1 & 1.96{$\pm 0.1$} & 104{$\pm 9$}	  & 147 &  &  \\ 
       	& 1.5{$\pm 0.1$}	& 142{$\pm 18$}	  & 93 &  &  \\ 
	& 1.7{$\pm 0.15$}	& 125{$\pm 21$}	  & 118 &  &  \\ 
	& 1.29{$\pm 0.08$}	& 168{$\pm 20$}	  & 72 &  &  \\ 
	& 1.81{$\pm 0.07$}	& 116{$\pm 8$}	  & 137 &  &  \\ 
NGC1313 source B & 1.47{$\pm 0.08$} & 110{$\pm 12$}	& 50 &  &  \\ 
       	         & 1.07{$\pm 0.07$} & $129_{-16}^{+19}$	& 20 &  &  \\ 
M81 source X6 & 1.59{$\pm 0.09$} & 72{$\pm 7$}	   & 30 &  &  \\ 
       	& 1.29{$\pm 0.13$}	& $89_{-15}^{+20}$ & 20 &  &  \\ 
\enddata 
\end{deluxetable} 



\end{document}